\documentclass{ceab}  

\usepackage{epsfig}    
\usepackage{graphicx}   

\usepackage{ceabbib}     

\usepackage[T1]{fontenc}

\begin{document}

\title{DEVELOPMENT AND MORPHOLOGY OF LEADING-FOLLOWING PARTS OF SUNSPOT GROUPS}

\author{Murak\"ozy J., T. Baranyi and A. Ludm\'any
\vspace{2mm}\\
\it Heliophysical Observatory, Hungarian Academy of Sciences, \\
\it H-4010 Debrecen, P.O.Box 30. Hungary}

\maketitle

\begin{abstract}

The detailed sunspot catalogues, the DPD and SDD allow to study the leading and following parts of sunspot groups separately. We examine the equilibrium distance of the two parts, the speed of removal, the asymmetry of compactness and the area growth. The distributions of positive and negative tilts of sunspot groups are also examined. 

\end{abstract}


\def\gore{sunspot group structure}

\section{Introduction}

The internal processes and morphological properties of solar active regions have been mostly subjects of case studies because of the earlier lack of detailed long term databases. Several internal features of sunspot groups are of diagnostic importance about the development of active regions, their interaction with the surrounding velocity fields or the advancement of the dynamo mechanism. The investigation of these topics became possible with the appearance of new photospheric databases. The present report lists some of these new topics and presents some preliminary results. The more detailed analysis of these phenomena is under way and will be published in more extended works.

\section{The observational material}

The observational data are taken from the SOHO/MDI - Debrecen sunspot Data (SDD), (Gy\H ori et al, 2011). This material is more detailed than the classic Greenwich Photoheliographic Results (GPR) and its continuation, the Debrecen Photoheliographic Data (DPD). These latter sunspot catalogues are suitable and indispensable for long-term studies of the solar activity, the solar cycles and dynamo, however, they are much less suitable to investigate all internal finer details and processes within the sunspot groups. This is partly because of the daily sampling, the observations are taken on the ground. On the other hand, the internal structure is mostly determined by the magnetic polarity relations and these data are missing from the GPR and DPD.

The SDD exploits the wealth of data provided by the SOHO/MDI instrument. The sunspot data are produced on an hourly basis, they comprise data of position, umbra-penumbra area and magnetic field for all observable sunspots and sunspot groups. Currently this data structure allows the highest resolution in both space and time. 

\begin{figure}[h]
  \begin{center}
   \epsfig{file=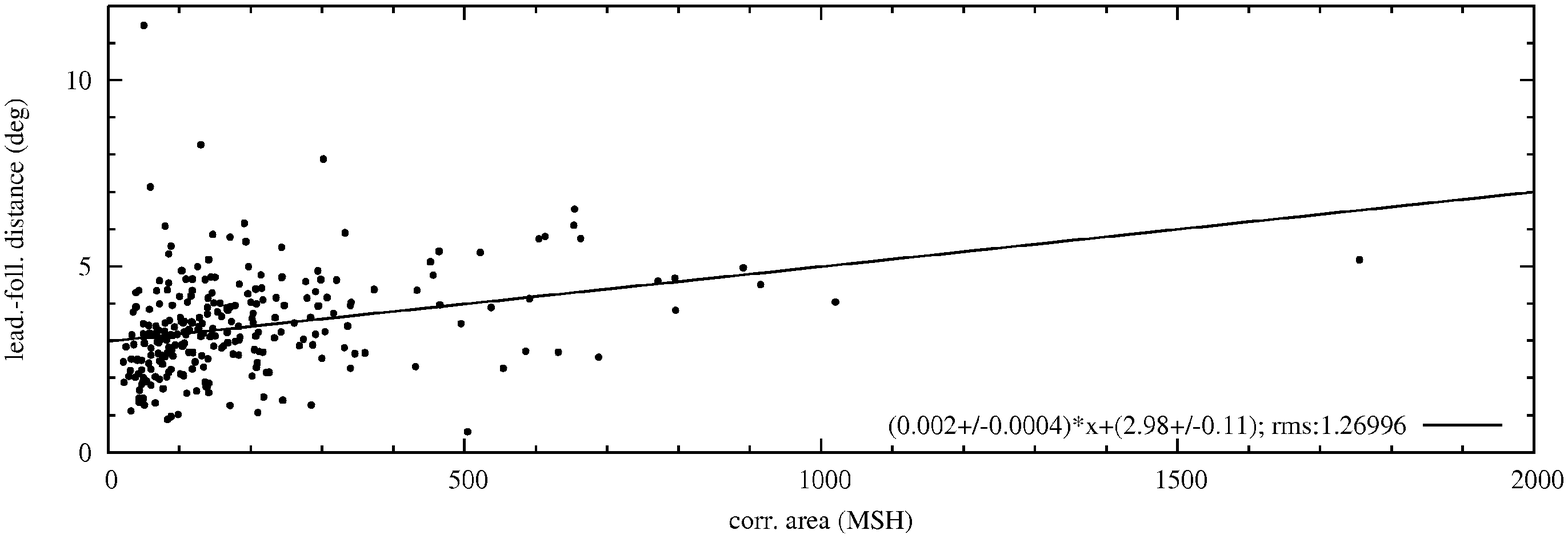,width=12cm}
   \epsfig{file=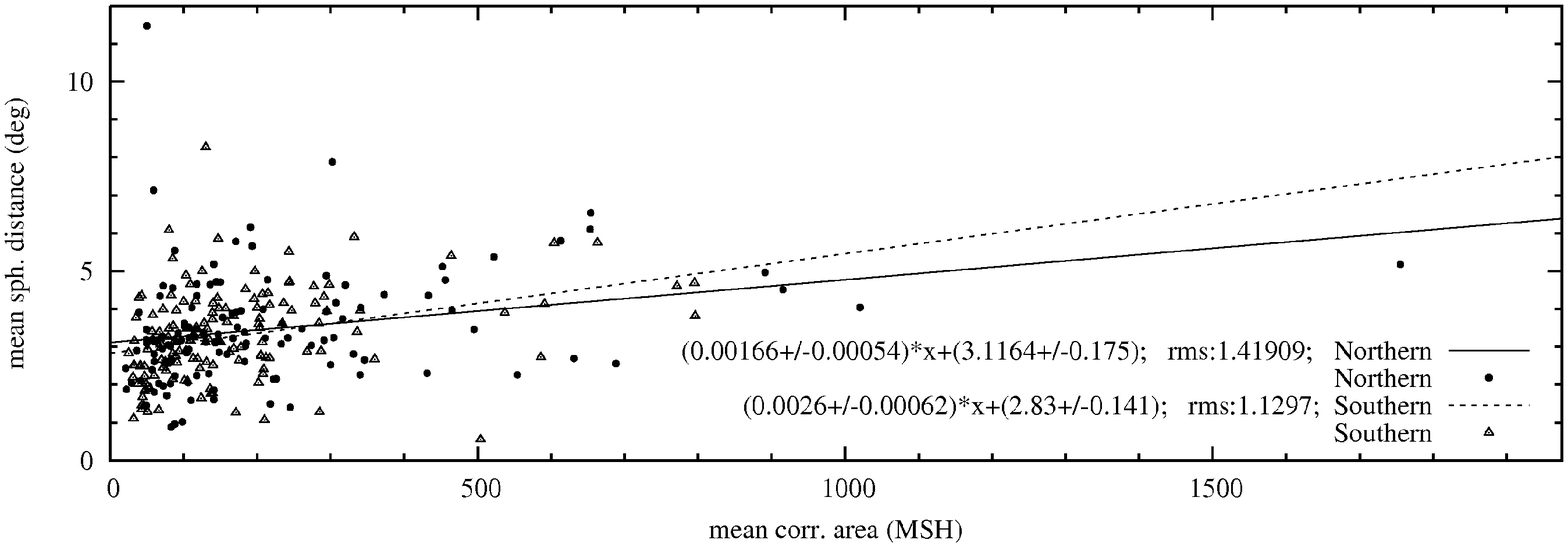,width=12cm}
  \end{center}
  \caption{Upper panel: distances between leading and following polarity regions depending on the total areas of sunspot groups. Lower panel: the same diagram as in the upper panel for the two hemispheres separately.}
\end{figure} 

\section{Distance of leading-following subgroups}

The polarity distance of emerging bipolar active regions develops more or less parallelly with the sunspot area and it reaches a quasi-equilibrium state at the maximum phase of the active region. This equilibrium exists between the magnetic tension of the active region flux ropes and mechanical impacts on the emerged flux ropes. The upper panel of Figure 1 shows the relationship between the total areas of sunspot groups and the distances between the centers of weight of their leading and following parts at the time of the largest total umbral area of the sunspot group. To ensure that in fact the time of maximum is considered only those sunspot groups are taken into account whose entire life-span was observed on the visible disc. This means a statistical sample of 272 active regions. The relationship is linear but fairly weak. This may indicate that the magnetic tension only plays a weak role in the formation of the active region size. The lower panel distinguishes the northern and southern hemispheres. Apparently there is no significant difference between the two hemispheres. 

\section{Compactness of sunspot groups}

The emerged active regions usually exhibit a specific asymmetry: the leading and following subgroups are not equally compact. The left panel of Figure 2 shows the relationship between the leading-following asymmetry indexes of spot numbers and mean spot areas. The data are computed at the time of the greatest total sunspot area of the active regions. The majority of the active regions belong to the upper left quarter of the diagram. This means that the typical configuration is characterized by smaller number and larger mean area of spots in the leading part. The right panel of Figure 2 compares how dispersed are the leading-following parts. 

\begin{figure}[ht]
\begin{center}
   \epsfig{file=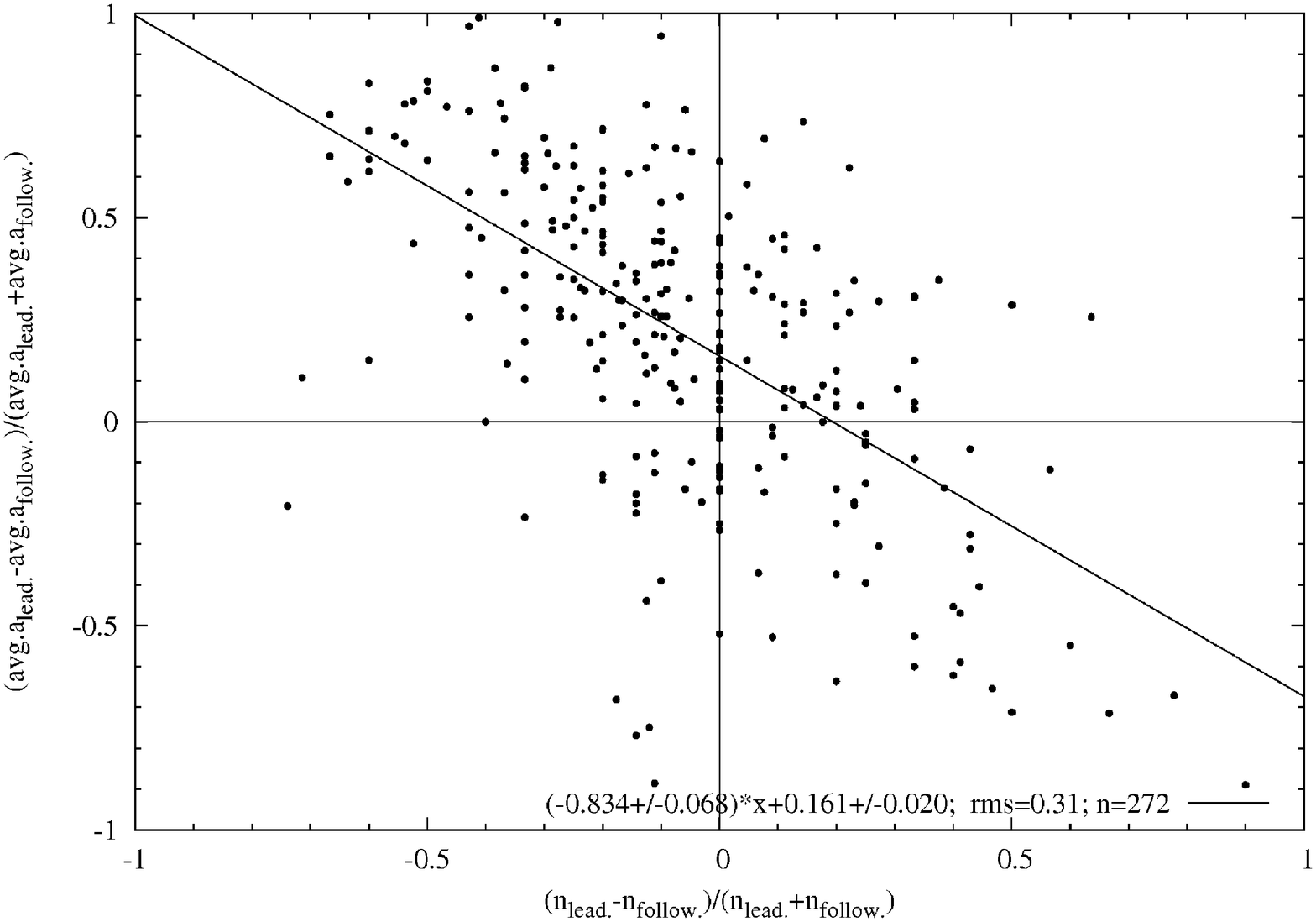,width=6cm}
   \epsfig{file=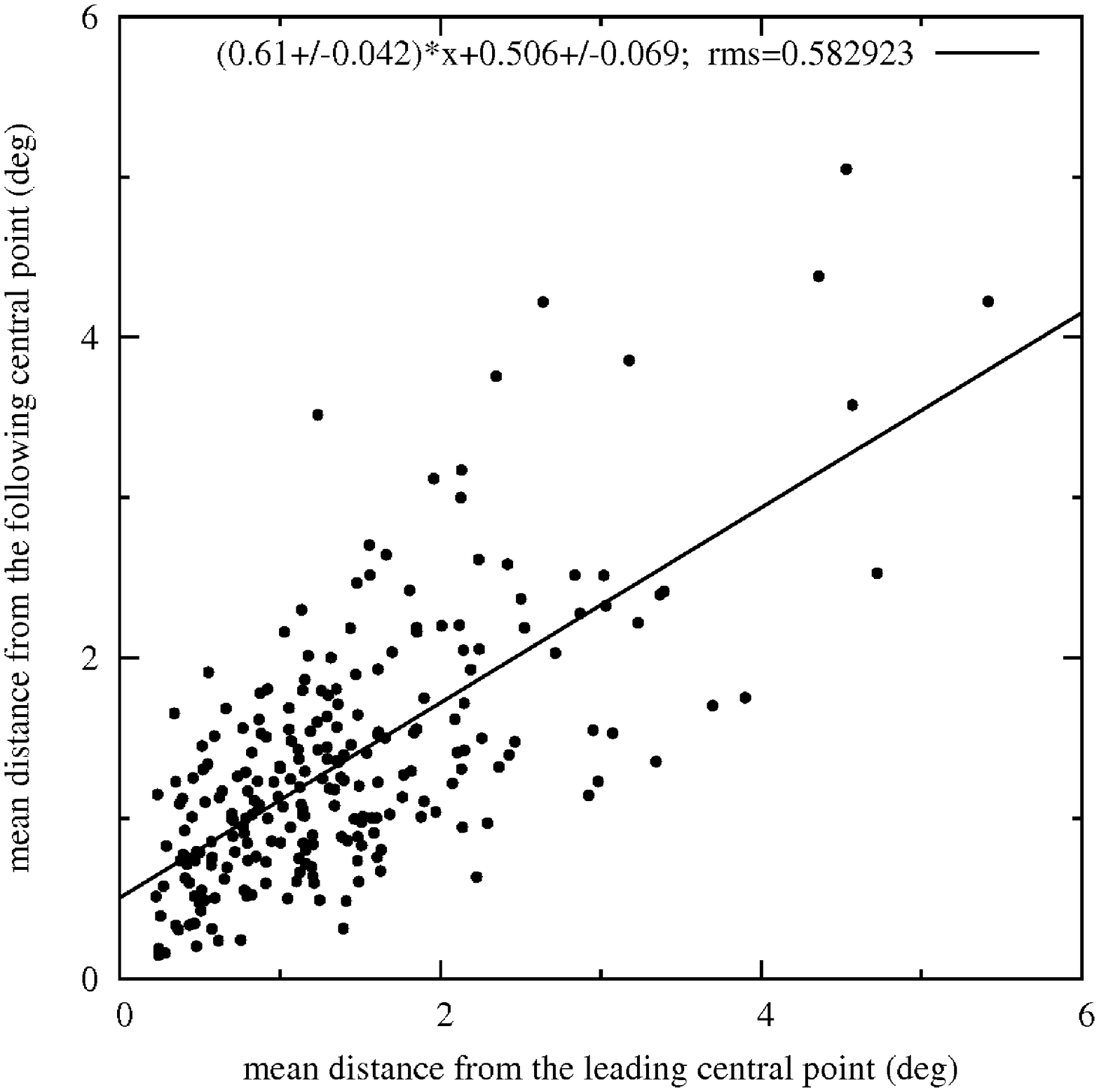,width=6cm}
\end{center}
  \caption{Left panel: relation between leading/following asymmetry indexes of numbers and mean areas of sunspots in sunspot groups. Right panel: comparison of the dispersednesses of leading and following subgroups, i.e. the mean distances of spots from the centers of weight in the leading and following parts.}
\end{figure} 

\section{Development of sunspot groups}

The emergence and decay of magnetic fields are important characteristics of the active region dynamics. They are related to different physical processes: the emergence is driven by buoyancy whereas the decay is the consequence of an erosion process caused by the turbulent environment of the flux ropes (Petrovay and van Driel-Gesztelyi, 1997; Petrovay et al., 1999). Nevertheless, these processes can be mixed during the development of the active region. The most recent relevant publication of Hathaway and Choudhary (2008) only refers to the development curve of the total area of a sunspot group from the GPR with daily resolution. The SDD opened the possibility to investigate the heading and trailing parts separately in hourly resolution. Two active regions were selected to demonstrate the new possibilities. The selection was motivated by the aim to have area development curves with reasonably complete increasing and decreasing phases. Development profiles of two active regions are plotted in Figure 3 describing the history of active regions NOAA 10988 and NOAA 10311. The heading and trailing parts are plotted separately along with the separation of their centers of weight. The fitted asymmetric gaussian is the following:

\begin{equation}
      f(t)=H\cdot exp {(-\frac{(t-M)^2}{D(1+At)})} 
         \label{asymcurve} 
\end{equation}

where H and M are the value and position of maximum, D and A determine the width and asymmetry respectively. 

The detailed investigation of the emergence/decay process will need an extended sampling but some properties can be noticed from these two examples. The heading part is larger, it develops more rapidly but it reaches the maximum somewhat later that the trailing part in both cases. 

\begin{figure}[h]
  \begin{center}
   \epsfig{file=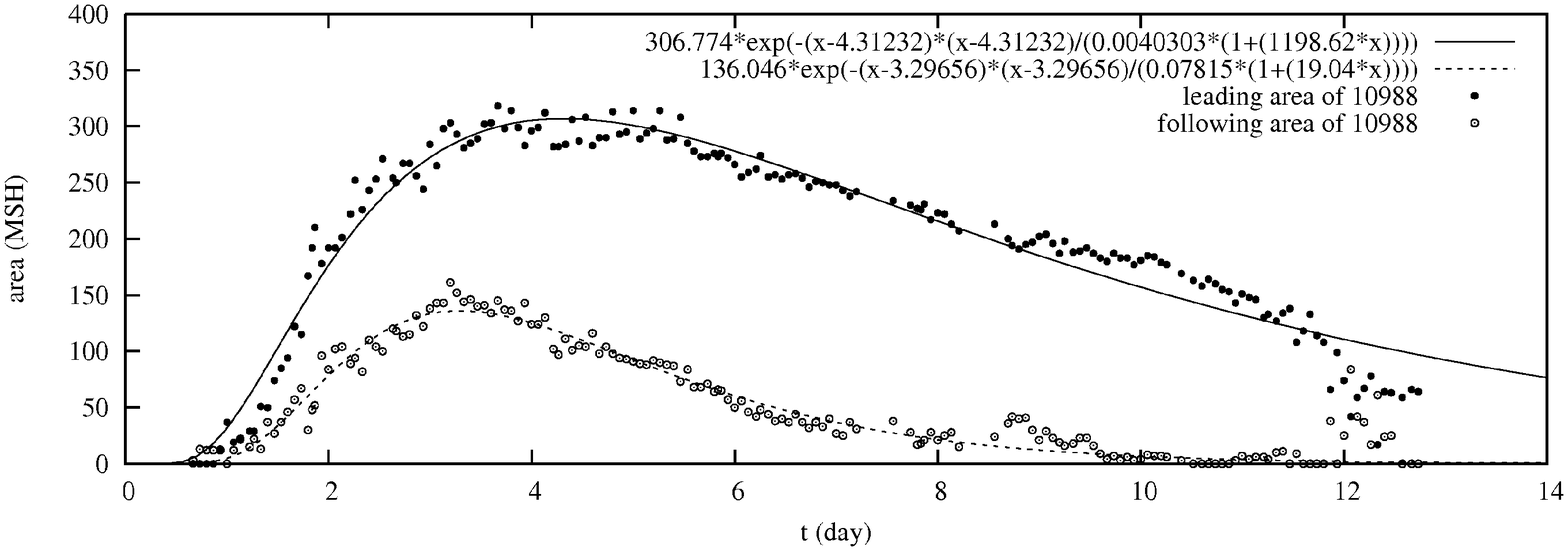,width=12cm}
   \epsfig{file=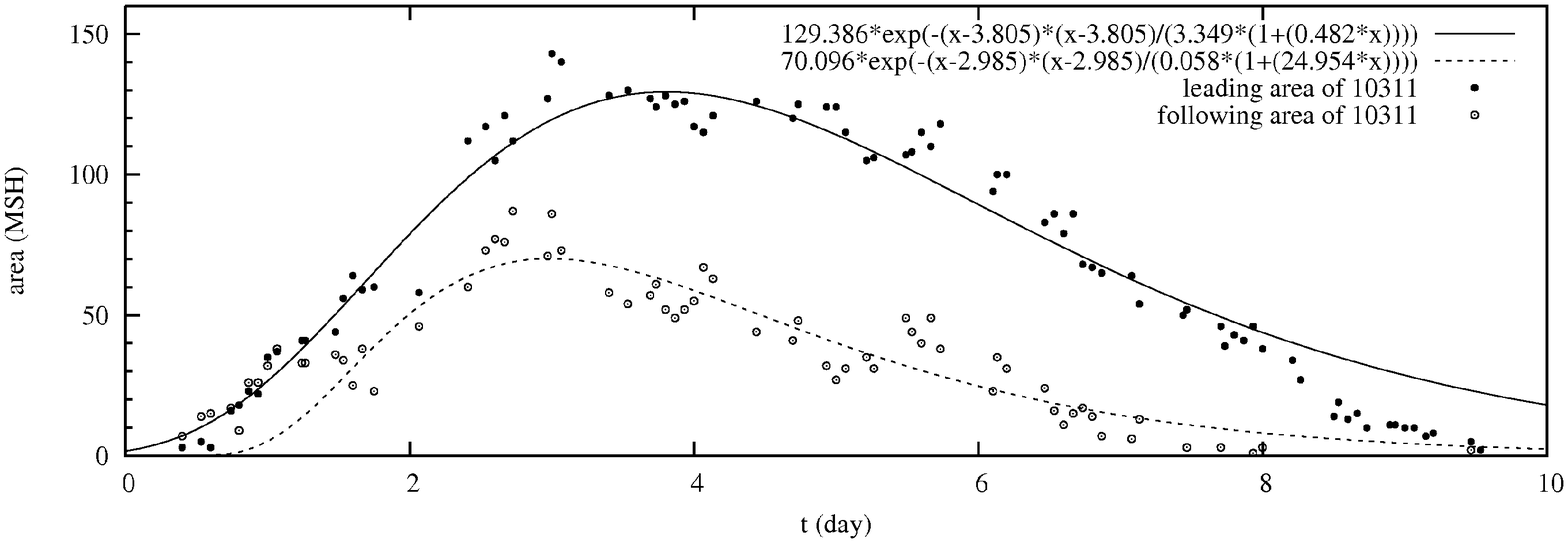,width=12cm}
   \end{center}
  \caption{Development of the active regions NOAA 10988 and NOAA 10311. The leading and trailing parts are plotted separately and an asymmetric Gauss function is fitted to the points.  }
\end{figure} 

\section{Tilts of sunspot groups}

The magnetic axes of the active regions (the axis is the line connecting the centers of weight of the areas of opposite polarities) have certain tilts with respect to the azimuthal direction, the leading part tends to be closer to the equator than the following part, this is considered to be positive tilt in both hemispheres. This property is known as the Joy's law. The underlying process is the winding up of the poloidal field to toroidal form, the process has been theoretically investigated by D'Silva and Choudhuri (1993) and Caligari et al. (1995). It should be noted, however, that the number of negative tilts cannot be neglected. The present report investigates the rate of positive/negative tilts. 

\begin{figure}[h]
  \begin{center}
   \epsfig{file=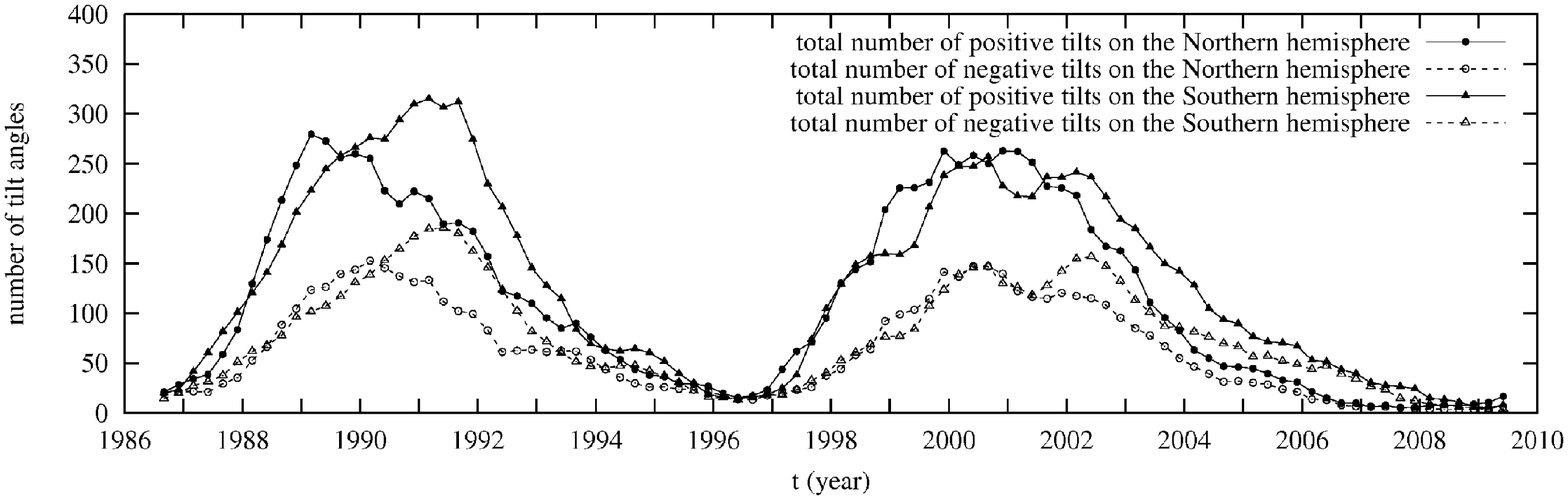,width=12cm}
   \epsfig{file=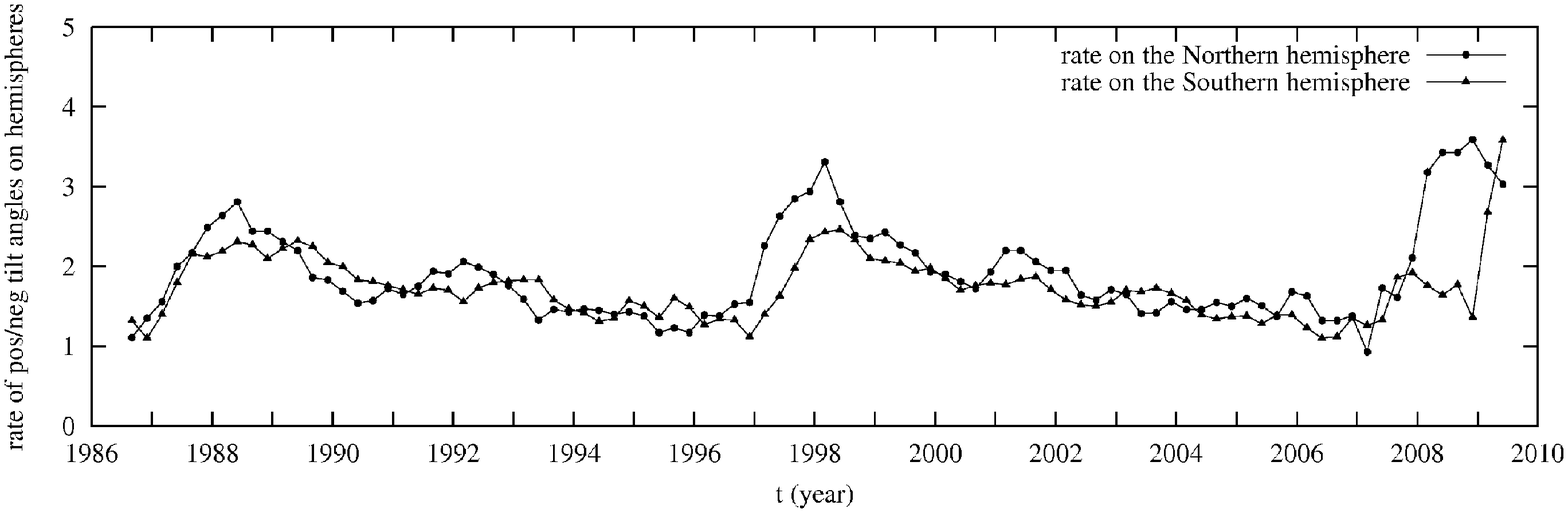,width=12cm}
  \end{center}
  \caption{Upper panel: positive and negative tilts in the northern and southern hemispheres in cycles 22-23. Lower panel: the ratio of positive/negative tilts in cycles 22-23.}
\end{figure} 

The DPD and SDD catalogues also contain the tilt angles. The upper panel of Figure 4 shows the numbers of positive and negative tilts at the times of the maximum areas of sunspot groups in both hemispheres in cycles 22-23. The comparison of the curves of positive and negative tilts shows that in the developing phase the majority of positive tilts is significant, its number is higher than that of the negative tilts by a factor of 2 or more, but after the maximum the difference gradually disappears and at the end of the cycle the two directions are similarly frequent. The reason is obviously the gradual approach of the toroidal field to the azimuthal direction and at the end of the cycle the tilts are randomly varying between positive and negative values. This is demonstrated in the lower panel of Figure 4. The ratios of positive/negative tilt angles are plotted for cycles 22-23 and the beginning of cycle 24 for the two hemispheres separately. In all cases the positive/negative ratios are higher at the beginnings of the cycles and they are diminishing until the ends of cycles. 

To have an impression about the spatial and temporal distribution of tilts their positive and negative values have been plotted separately in the Schwabe-diagram, see Figure 5. The size of the pixels are: a quarter of a year and $5^{\circ}$ in latitude, the data are averaged within these domains. The tilt data of DPD are determined for two objects, both for the umbrae and penumbrae and the diagram only takes into account those cases when the difference of these two data was smaller than $5^{\circ}$, this restriction eliminates the most ambiguous cases. It is conspicuous that the highest mean values are found at the border of the butterfly diagram for both negative and positive angles, in the central region of the diagram the moderate mean positive tilts are overwhelming.

\begin{figure}[h]
  \begin{center}
   \epsfig{file=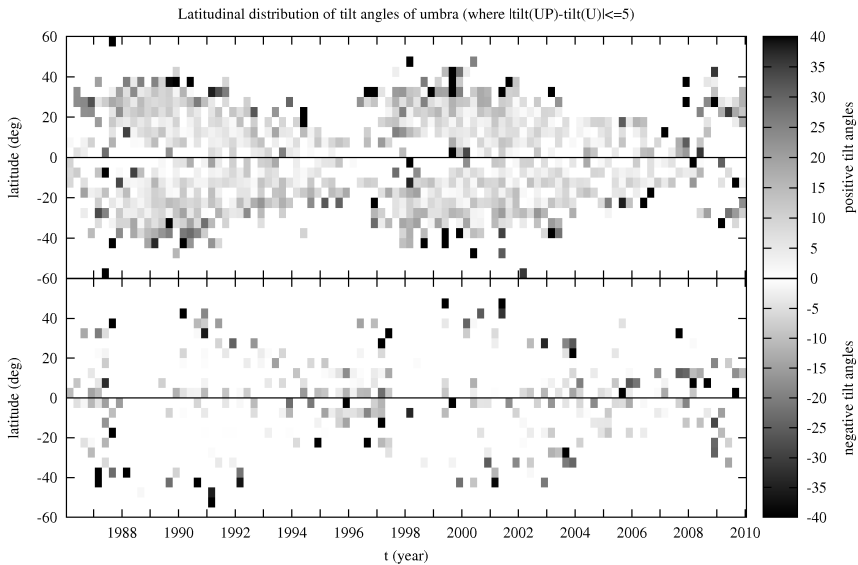,width=12cm}
  \end{center}
  \caption{Distribution of active region tilt values in the butterfly diagrams of cycles 22-23. The upper and lower panels show the positive and negative angles respectively.}
\end{figure}

\section{Discussion}

The presented features are preliminary results of a series of ongoing investigations. The study of these topics became possible by the new detailed photospheric datasets. The leading-following differences have diagnostic importance about the dynamics of active regions. The role of polarity separation is studied by D'Silva and Choudhuri (1993), the compactness asymmetry is analysed by Fan et al. (1993, 1994), the sunspot decay is studied by Petrovay et al. (1999). The best studied feature is the behaviour of active region tilts, an important ingredient of the flux transport dynamo models. Recently, Dasi-Espuig et al. (2010) reports evidences for the role of the tilts in the preparation of the next cycle. The forthcoming extension of the above examinations will provide more empirical facts to these investigations.

\section*{Acknowledgements} 
The research leading to these results has received funding from the European Community's Seventh Framework Programme (FP7/2007-2013) under grant agreement No. 218816.


\section*{References}
\begin{itemize}
\small
\itemsep -2pt
\itemindent -20pt

\item[] Caligari, P., Moreno-Insertis, F., Sch\"ussler, M.,  1995, {\it \apj}, 441, 886.
\item[] Dasi-Espuig, M.; Solanki, S. K.; Krivova, N. A.; Cameron, R.; Pe\~nuela, T., 2010, {\it \aap}, 518, 7.
\item[] D'Silva, S.; Choudhuri, A. R., 1993, {\it \aap}, 272, 621-633.
\item[] Fan, Y., Fisher, G. H. \& DeLuca, E. E., 1993, {\it \apj}, 405, 390-401.
\item[] Fan, Y., Fisher, G. H. \& McClymont, A. N., 1994, {\it \apj}, 436, 907-928.
\item[] Gy\H ori, L., Baranyi, T., \& Ludm\'any, A. 2011, IAU Symp. 273, 403. \\ see: http://fenyi.solarobs.unideb.hu/SDD/SDD.html
\item[] Hathaway, D. \& Choudhary, D. P., 2008, {\it \solphys}, 250, 269-278.
\item[] Petrovay, K.; van Driel-Gesztelyi, L., 1997, {\it \solphys}, 176, 249-266.
\item[] Petrovay, K.; Mart\'{i}nez Pillet, V.; van Driel-Gesztelyi, L., 1999,  {\it \solphys}, 188, 315-330.
\item[] 

\end{itemize}

\end{document}